\documentstyle[sprocl]{article}

\newcommand{\plb}[2]{{\em Phys. Lett.}              {\bf #1B}, #2 }
\newcommand{\npb}[2]{{\em Nucl. Phys.}              {\bf B#1}, #2 }
\newcommand{\npp}[2]{{\em Nucl. Phys. Proc. Suppl.} {\bf  #1}, #2 }

\newcommand{\prd}[2]{{\em Phys. Rev.}               {\bf D#1}, #2 }
\newcommand{\prl}[2]{{\em Phys. Rev. Lett.}         {\bf  #1}, #2 }
\newcommand{\zpc}[2]{{\em Z. Phys.}                 {\bf C#1}, #2 }
\newcommand{\sci}[2]{{\em Science}                  {\bf  #1}, #2 }
\newcommand{\jpb}[2]{{\em J. Phys.}                 {\bf B#1}, #2 }

\newcommand{\epj}[2]{{\em Eur. Phys. J.}            {\bf C#1}, #2 }
\newcommand{\con}[2]{                               {\bf  #1}, #2 }
\newcommand{\etal}{{\em et al.}}
\newcommand{\ibid}{{\em ibid.}}
\newcommand{\col}{Collaboration}

\newcommand{\be}{\begin{equation}}
\newcommand{\ee}{\end{equation}}
\newcommand{\ba}{\begin{array}}
\newcommand{\ea}{\end{array}}
\newcommand{\ms}{{\overline{\rm MS}}}
\newcommand{\lsim}{\buildrel < \over {_\sim}}
\newcommand{\gsim}{\buildrel > \over {_\sim}}

\def\be{\begin{equation}}
\def\ee{\end{equation}}
\def\bea{\begin{eqnarray}}
\def\eea{\end{eqnarray}}

\begin{document}

\hfill{UPR--0816--T}
\vspace{15pt} 

\title{STATUS OF THE STANDARD MODEL\footnote{To appear in the 
Proceedings of the 5th International WEIN Symposium: 
A Conference on Physics Beyond the Standard Model (WEIN 98),
Santa Fe, NM, June 14--21, 1998.}
}

\author{J. ERLER, P. LANGACKER}

\address{Department of Physics and Astronomy,
University of Pennsylvania, \\
Philadelphia, PA 19104-6396, USA \\
E-mail:  erler@langacker.hep.upenn.edu; pgl@langacker.hep.upenn.edu}


\maketitle\abstracts{ We review the experimental and theoretical status of the
standard electroweak theory, and its fundamental parameters. We obtain the 
global best fit result for the Standard Model Higgs boson of 
$M_H = 107^{+67}_{-45}$ GeV, and find the 95\% upper limit of 255 GeV.  
Parameters describing physics beyond the Standard Model are discussed as well. 
Particular emphasis is given to implications for supersymmetric extensions 
of the Standard Model.}

\section{The $Z$ and the Weak Neutral Current}
Weak neutral currents are a primary prediction and a direct test
of electroweak unification.  They were discovered in 1973 by the Gargamelle
experiment using the Proton Synchrotron (PS) at CERN~\cite{Hasert73}, and
confirmed by the HPWF detector at Fermilab~\cite{Benvenuti74}.
Subsequently, neutrino-nucleon and neutrino-electron scattering experiments
improved to the per cent level, testing the weak interaction quantitatively. 
Electron-deuteron and electron-positron scattering, as well
as atomic parity violation experiments, are sensitive to weak-electromagnetic 
interference effects, and were crucial for the confirmation of the electroweak 
Standard Model (SM). The $W$ and $Z$ bosons were finally discovered directly
by the UA1~\cite{Arnison83} and UA2~\cite{Banner83} Collaborations at the 
Super Proton Synchrotron (SPS) at CERN in 1982 and 1983, respectively. For a 
description of the history we refer the reader to Ref.~\cite{Mann94}.

With the basic structure of the SM established, CERN's Large Electron Positron
accelerator (LEP) and SLAC's Stanford Linear Collider (SLC) were designed 
to test it on the quantum level. With these machines it was possible to 
determine many $Z$ properties with per mille accuracy, including the 
outstanding measurement of the $Z$ mass, $M_Z$, at LEP 1 with a relative 
precision of 2 parts in $10^5$. Run I of Fermilab's Tevatron (CDF and D\O ) 
and the second phase of LEP (ALEPH, DELPHI, L3, and OPAL) contribute per mille
determinations of the $W$ mass, $M_W$.  With the mass of the top quark, $m_t$,
as determined at the Tevatron, $M_W$ and other high precision observables can 
also be calculated within the SM in the framework of a quantum field theory. 
The agreement with the measurements establishes the SM as a spontaneously 
broken renormalizable gauge theory, and verifies the gauge group and 
representations. It predicts at least one extra state, the Higgs boson, with 
a mass, $M_H$, below 1 TeV (from triviality bounds and for reasons of 
perturbativity). Combining all direct and indirect data in a likelihood fit, 
it is possible to extract more precise information on $M_H$ leading to upper 
bounds of at most a few hundred GeV. We will review SM parameter estimation 
in Section~\ref{sm}.

The high accuracy of theory and experiment allows severe constraints
on possible TeV scale physics, such as unification or compositeness. 
For example, the ideas of technicolor and non-supersymmetric Grand Unified 
Theories (GUT's) are strongly disfavored. On the other hand, supersymmetric 
unification, as generically predicted by string theories, is supported by the
observed approximate gauge coupling unification at an energy slightly
below the Planck scale. Constraints on parameters describing physics
beyond the SM will be reviewed in Section~\ref{np}.

\section{New Data}
\label{newdata}
\subsection{$Z$ Pole Physics}
\label{zpolephys}
The most important input into precision tests of electroweak theory continues 
to come from the $Z$ factories LEP 1~\cite{Abbaneo97} and SLC~\cite{Baird98}. 
The vanguard of the physics program at LEP~1 with about 20 million recorded $Z$
events is the analysis of the $Z$ lineshape. Its parameters are $M_Z$, the 
total $Z$ width, $\Gamma_Z$, the hadronic peak cross section, 
$\sigma_{\rm had}$, and the ratios of hadronic to leptonic decay widths, 
$R_\ell = {\Gamma({\rm had})\over \Gamma(\ell^+\ell^-)}$, where $\ell = e$, 
$\mu$, or $\tau$. They are determined in a common fit with the leptonic 
forward-backward (FB) asymmetries, $A_{FB} (\ell) = {3\over 4} A_e A_\ell$. 
Here 
\be
  A_f = {2 v_f a_f\over v_f^2 + a_f^2}
\ee
is a parameter for fermion $f$, defined in terms of the vector 
($v_f = I_{3,f} - 2 Q_f \sin^2 \theta_f^{\rm eff}$) and axial-vector 
($a_f = I_{3,f}$) $Zf\bar{f}$ coupling; $Q_f$ and $I_{3,f}$ are the electric 
charge and third component of isospin, respectively, and 
$\sin^2 \theta_f^{\rm eff} \equiv \bar{s}^2_f$ is an effective mixing angle.

An average of about 73\% polarization of the electron beam at the SLC allows 
for a set of competitive and complementary measurements with a much smaller 
number of $Z$'s ($\gsim 500,000$). In particular, the left-right (LR) cross 
section asymmetry, $A_{LR} = A_e$, represents the most precise determination 
of the weak mixing angle by a single experiment (SLD)~\cite{Baird98}. 
Mixed FB-LR asymmetries, $A^{FB}_{LR} (f) = {3\over 4} A_f$, single out the 
final state coupling of the $Z$ boson. This is done for leptons~\cite{Baird98},
$s$ quarks~\cite{Abe98}, as well as $b$ and $c$ quarks. 

The results for $A^{FB}_{LR} (\tau)$ (and $A^{FB}_{LR} (e)$) can directly be 
compared with the LEP results~\cite{Karlen98} on the final state $\tau$ 
polarization, ${\cal P}_\tau$ (and its angular distribution, 
${\cal P}^{FB}_\tau$). For several years there has been an experimental 
discrepancy at the $2 \sigma$ level between $A_\ell$ from LEP and the 
SLC~\cite{Erler98}. With the 1997/98 high statistics run\footnote{There is 
still some discrepancy in $A_{LR}$ from the SM prediction. This is mostly
from upward fluctuations in the 1993 and 1996 data. The preliminary results 
from 1997, $A_{LR} = 0.1475 \pm 0.0042 \pm 0.0016$, and 1998,
$A_{LR} = 0.1487 \pm 0.0031 \pm 0.0017$, are in excellent agreement.} at the 
SLC, and a revised value for ${\cal P}^{FB}_\tau$, the two determinations are 
now consistent with each other,
\be \ba{l}
\label{aell}
  A_\ell ({\rm LEP}) = 0.1470 \pm 0.0027, \\
  A_\ell ({\rm SLD}) = 0.1503 \pm 0.0023.
\ea \ee
The LEP value is from $A_{FB} (\ell)$ and the $\tau$ polarization measurements,
while the SLD value is from $A_{LR}$ and $A^{FB}_{LR} (\ell)$. The data are 
consistent with lepton universality, which is assumed here. There remains, 
however, a $2.5 \sigma$ discrepancy between the two most precise determinations
of $\bar{s}^2_\ell$, namely $A_{LR}$ and $A_{FB} (b)$ (assuming no new physics
in $A_b$). 

Of particular interest are the results on the heavy flavor 
sector~\cite{Karlen98} including 
$R_q = {\Gamma (q\bar{q}) \over \Gamma ({\rm had})}$, $A_{FB} (q)$, and 
$A^{FB}_{LR} (q)$, with $q = b$ or $c$. In addition, results have been quoted 
on $A_{FB} (s)$~\cite{Abreu95,Ackerstaff98} and 
$R_{d,s}/(R_d + R_u + R_s)$~\cite{Ackerstaff98}. There is a theoretical 
prejudice that the third family is the one which is most likely affected by 
new physics. Interestingly, the heavy flavor sector has always shown the 
largest deviations from the SM. E.g., $R_b$ deviated at times by almost 
$4 \sigma$. Now, however, $R_b$ is in good agreement with the SM, and thus 
puts strong constraints on many types of new physics. At present, there is some
discrepancy in $A^{FB}_{LR} (b) = {3\over 4} A_b$, and 
$A_{FB} (b) = {3\over 4} A_e A_b$, both at the $2 \sigma$ level. Using 
the average of Eqs.~(\ref{aell}), $A_\ell = 0.1489 \pm 0.0018$, both can be 
interpreted as measurements of $A_b$. From $A_{FB} (b)$ one would obtain 
$A_b = 0.887 \pm 0.022$, and the combination with 
$A^{FB}_{LR} (b) = {3\over 4} (0.867 \pm 0.035)$ would yield 
$A_b = 0.881 \pm 0.019$, which is almost $3 \sigma$ below the SM prediction. 
Alternatively, one could use $A_\ell ({\rm LEP})$ above (which is closer to the
SM prediction) to determine $A_b ({\rm LEP}) = 0.898 \pm 0.025$, and 
$A_b = 0.888 \pm 0.020$ after combination with $A^{FB}_{LR} (b)$, i.e., still 
a $2.3 \sigma$ discrepancy. In order to explain this 5--6\% deviation in $A_b$
in terms of new physics in loops, a 25--30\% radiative correction to 
$\hat\kappa_b$ defined through $\bar{s}^2_b = \hat\kappa_b\sin^2\hat\theta_\ms$
would be needed. Only a new type of physics which couples at the tree level 
preferentially to the third generation~\cite{Erler95}, and which does not 
contradict $R_b$ (including the off-peak measurements by 
DELPHI~\cite{Abreu96}), can conceivably account for a low $A_b$. 
Given this and that none of the observables deviates by $2 \sigma$ or more, 
we can presently conclude that there is no compelling evidence for new physics 
in the precision data. 

LEP also quotes a value for the hadronic charge asymmetry, $Q_{FB}$,
representing an additional determination of $\bar{s}_\ell^2 (A_{FB} (q))$.
Similarly, SLD measures the hadronic charge flow asymmetry~\cite{Abe97}, 
$A_Q = A_e$, which is basically given by the ratio of (weighted) FB and 
LR--FB asymmetries.

\begin{table}[h]
\caption{$Z$ pole precision observables from LEP and the SLC.
Shown are the experimental results, the SM predictions, and the pulls.
The SM errors are from the uncertainties in $M_Z$, $\ln M_H$, $m_t$, 
$\alpha (M_Z)$, and $\alpha_s$. They have been treated as Gaussian 
and their correlations have been taken into account.} 
\label{zpole}
\vspace{0.2cm}
\begin{center}
\footnotesize
\begin{tabular}{|l|c|c|c|r|}
\hline Quantity & Group(s) & Value & Standard Model & pull \\ 
\hline
$M_Z$ \hspace{14pt}      [GeV]&     LEP     &$ 91.1867 \pm 0.0021 $&$ 91.1865 \pm 0.0021 $&$ 0.1$ \\
$\Gamma_Z$ \hspace{17pt} [GeV]&     LEP     &$  2.4939 \pm 0.0024 $&$  2.4957 \pm 0.0017 $&$-0.8$ \\
$\Gamma({\rm had})$\hspace{8pt}[GeV]&  LEP  &$  1.7423 \pm 0.0023 $&$  1.7424 \pm 0.0016 $&  ---  \\
$\Gamma({\rm inv})$\hspace{11pt}[MeV]& LEP  &$500.1    \pm 1.9    $&$501.6    \pm 0.2    $&  ---  \\
$\Gamma({\ell^+\ell^-})$ [MeV]&     LEP     &$ 83.90   \pm 0.10   $&$ 83.98   \pm 0.03   $&  ---  \\
$\sigma_{\rm had}$ \hspace{12pt}      [nb] &     LEP     &$ 41.491  \pm 0.058  $&$ 41.473  \pm 0.015  $&$ 0.3$ \\
$R_e$                         &     LEP     &$ 20.783  \pm 0.052  $&$ 20.748  \pm 0.019  $&$ 0.7$ \\
$R_\mu$                       &     LEP     &$ 20.789  \pm 0.034  $&$ 20.749  \pm 0.019  $&$ 1.2$ \\
$R_\tau$                      &     LEP     &$ 20.764  \pm 0.045  $&$ 20.794  \pm 0.019  $&$-0.7$ \\
$A_{FB} (e)$                  &     LEP     &$  0.0153 \pm 0.0025 $&$  0.0161 \pm 0.0003 $&$-0.3$ \\
$A_{FB} (\mu)$                &     LEP     &$  0.0164 \pm 0.0013 $&$                    $&$ 0.2$ \\
$A_{FB} (\tau)$               &     LEP     &$  0.0183 \pm 0.0017 $&$                    $&$ 1.3$ \\
\hline
$R_b$                         &  LEP + SLD  &$  0.21656\pm 0.00074$&$  0.2158 \pm 0.0002 $&$ 1.0$ \\
$R_c$                         &  LEP + SLD  &$  0.1735 \pm 0.0044 $&$  0.1723 \pm 0.0001 $&$ 0.3$ \\
$R_{s,d}/R_{(d+u+s)}$         &     OPAL    &$  0.371  \pm 0.023  $&$  0.3592 \pm 0.0001 $&$ 0.5$ \\
$A_{FB} (b)$                  &     LEP     &$  0.0990 \pm 0.0021 $&$  0.1028 \pm 0.0010 $&$-1.8$ \\
$A_{FB} (c)$                  &     LEP     &$  0.0709 \pm 0.0044 $&$  0.0734 \pm 0.0008 $&$-0.6$ \\
$A_{FB} (s)$                  &DELPHI + OPAL&$  0.101  \pm 0.015  $&$  0.1029 \pm 0.0010 $&$-0.1$ \\
$A_b$                         &     SLD     &$  0.867  \pm 0.035  $&$  0.9347 \pm 0.0001 $&$-1.9$ \\
$A_c$                         &     SLD     &$  0.647  \pm 0.040  $&$  0.6676 \pm 0.0006 $&$-0.5$ \\
$A_s$                         &     SLD     &$  0.82   \pm 0.12   $&$  0.9356 \pm 0.0001 $&$-1.0$ \\
\hline
$A_{LR}$ (hadrons)            &     SLD     &$  0.1510 \pm 0.0025 $&$  0.1466 \pm 0.0015 $&$ 1.8$ \\
$A_{LR}$ (leptons)            &     SLD     &$  0.1504 \pm 0.0072 $&$                    $&$ 0.5$ \\
$A_\mu$                       &     SLD     &$  0.120  \pm 0.019  $&$                    $&$-1.4$ \\
$A_\tau$                      &     SLD     &$  0.142  \pm 0.019  $&$                    $&$-0.2$ \\
$A_e (Q_{LR})$                &     SLD     &$  0.162  \pm 0.043  $&$                    $&$ 0.4$ \\
$A_\tau ({\cal P}_\tau)$      &     LEP     &$  0.1431 \pm 0.0045 $&$                    $&$-0.8$ \\
$A_e ({\cal P}_\tau)$         &     LEP     &$  0.1479 \pm 0.0051 $&$                    $&$ 0.3$ \\
$\bar{s}_\ell^2 (Q_{FB})$     &     LEP     &$  0.2321 \pm 0.0010 $&$  0.2316 \pm 0.0002 $&$ 0.5$ \\
\hline
\end{tabular}
\end{center}
\end{table}

The high precision $Z$ pole observables are summarized in Table~\ref{zpole}. 
Given for each observable is the experimental value with the total 
( = statistical + systematic, added in quadrature) error, the SM prediction, 
and the pull, i.e., the deviation from the SM normalized by the total error. 
$\Gamma({\rm had})$, $\Gamma({\rm inv})$, and $\Gamma({\ell^+\ell^-})$ are 
derived quantities and given for illustration only\footnote{The invisible 
width, $\Gamma({\rm inv})$, constrains the number, $N_\nu$, of standard 
neutrino flavors. A fit to all data with $N_\nu$ free yields 
$N_\nu = 2.992 \pm 0.011$.}. Very good agreement with the SM is observed. 
Only $A_{LR}$ and the two measurements sensitive to $A_b$ discussed above, 
show some deviation, but even those are below $2\sigma$.

\subsection{LEP 2 and Tevatron Results}
LEP 2~\cite{Karlen98} operating at and above the $W^+ W^-$ threshold, and the 
CDF~\cite{Dorigo98} and D\O\ \cite{Abbott98} \col s at the Tevatron, produce 
two completely independent determinations of $M_W$, with presently the same 
accuracy. They are in very good agreement with each other, and yield the world 
average (including the older result by the UA2 \col~\cite{Alitti92}),
\be
  M_W = 80.388 \pm 0.063 \mbox{ GeV}.
\ee

The determination of $m_t$ by CDF and D\O\ is dominated by the lepton + jet 
channels, which combine the merits of statistics (hadrons) and cleanliness 
(leptons). The dilepton channels (CDF and D\O ) and the all hadronic channel 
(CDF) add extra information with to some extent different systematic
uncertainties. The combined value is~\cite{Partridge98}
\be
  m_t = 173.8 \pm 3.2 \mbox{ (stat.)} \pm 3.9 \mbox{ (syst.)} \mbox{ GeV}.
\ee

Fermion pair production at LEP above the $Z$ resonance yields a number
of important tests and cross checks of the SM. For example, $\gamma$--$Z$ 
interference effects, which are suppressed at the $Z$ resonance, become more 
sizable at higher energies. The LEP \col s have performed additional 
``S-matrix'' fits, in which they allow these interference effects to differ 
from the SM expectations. They allow three extra off-resonance parameters, 
analogous and in addition to the on-resonance parameters $\sigma_{\rm had}$, 
$R_\ell$, and $A_{FB} (\ell)$. Including $M_Z$ and $\Gamma_Z$ and assuming 
family-universality this represents an eight parameter fit. Good agreement
with the SM is found, reflecting the fact that FB-asymmetry measurements
above the $Z$ peak are also in good agreement with SM expectations. 

The measurements at LEP 2 above the $W^+ W^-$ threshold and from D\O\ at the 
Tevatron are sensitive to triple-gauge-boson vertices. While there are a total 
of 14 independent couplings, one can use $SU(2)\times U(1)$ gauge invariance 
and LEP~1 constraints to reduce the number of triple gauge couplings to three. 
Each coupling is extracted from the data by setting the other two to zero
(the SM value). The results are~\cite{Karlen98}
\be
\ba{rcr}
   \Delta \kappa_\gamma &=&   0.13 \pm 0.14, \\
   \Delta g_1^Z         &=&   0.00 \pm 0.08, \\
         \lambda_\gamma &=& - 0.03 \pm 0.07.
\ea
\ee
        
\subsection{Low Energy Data}
Deep inelastic neutrino-hadron scattering (DIS) experiments played
an important role for the establishment of the SM, and now serve as
quantitative probes. This year the NuTeV \col~\cite{McFarland98} 
at Fermilab announced a very precise measurement of the 
Paschos-Wolfenstein~\cite{Paschos73} ratio,
\be
  R^- = {R^\nu - r R^{\bar{\nu}}\over 1 - r} = g_L^2 - g_R^2.
\ee
Here, $R^\nu$ ($R^{\bar{\nu}}$) is the ratio of neutral to charged current 
(anti)neutrino scattering cross sections, while $r$ is the ratio of 
charged current $\nu$ to $\bar{\nu}$ cross sections. $R^-$ has an
advantage over the more traditional DIS observable $R^\nu$ in that the effects
of scattering from sea quarks cancel in the difference of $\nu$ and
$\bar{\nu}$ cross sections. The actual measured quantity,
\be
  R^-_{\rm meas} = R^\nu_{\rm meas} - \alpha R^{\bar{\nu}}_{\rm meas},
\ee
is constructed to minimize uncertainties from the so-called
slow-rescaling parameters associated with the charm threshold, which have
been dominant in the past. $\alpha = 0.5136$ has been obtained by means of 
a Monte Carlo simulation. This method can only be applied when a high 
statistics $\bar{\nu}$ beam is available. Within the SM (but not beyond)
and after fixing\footnote{Note, that the $R^{\bar{\nu}}$ component
introduces a larger $m_t$ dependence compared to earlier DIS measurements,
such as from CCFR.} $m_t$ and $M_H$, the result, 
\be
  R^-_{\rm meas} = 0.2277 \pm 0.0022,
\ee
can be expressed as a measurement~\cite{McFarland98} of 
$M_W = 80.26 \pm 0.11$~GeV. This can be compared with the final result of the 
CCFR experiment~\cite{McFarland98A}, $M_W = 80.35 \pm 0.21$~GeV. In practical 
numerical implementations, especially in the presence of new physics, the 
actual combination of $Zq\bar{q}$ couplings should be used. In our analyses we
include earlier results by the CDHS~\cite{Blondel90} and CHARM~\cite{Allaby87}
\col s, as well, and take into account correlations induced by physics model 
uncertainties. For a recent update on the physics model parameters, see 
Ref.~\cite{Perrier95}.

\begin{table}[bht]
\caption{Non-$Z$ pole precision observables from the Tevatron, LEP 2, neutrino
scattering and APV. Shown are the experimental results, the SM predictions, and
the pulls. The second error after the experimental value, where given, is 
theoretical. The SM errors are from the inputs as in Table 1. The CHARM results
have been adjusted to CDHS conditions, and can be directly compared.}
\label{nonzpole}
\vspace{0.2cm}
\begin{center}
\footnotesize
\begin{tabular}{|l|c|c|c|r|}
\hline Quantity & Group(s) & Value & Standard Model & pull \\ 
\hline
$m_t$\hspace{8pt}[GeV]&Tevatron &$ 173.8    \pm 5.0               $&$ 171.4    \pm 4.8    $&$ 0.5$\\
$M_W$ [GeV]    & Tevatron + UA2 &$  80.404  \pm 0.087             $&$  80.362  \pm 0.023  $&$ 0.5$\\
$M_W$ [GeV]    &      LEP       &$  80.37   \pm 0.09              $&$                     $&$ 0.1$\\
\hline
$R^-$          &     NuTeV      &$   0.2277 \pm 0.0021 \pm 0.0007 $&$   0.2297 \pm 0.0003 $&$-0.9$\\
$R^\nu$        &     CCFR       &$   0.5820 \pm 0.0027 \pm 0.0031 $&$   0.5827 \pm 0.0005 $&$-0.2$\\
$R^\nu$        &     CDHS       &$   0.3096 \pm 0.0033 \pm 0.0028 $&$   0.3089 \pm 0.0003 $&$ 0.2$\\
$R^\nu$        &     CHARM      &$   0.3021 \pm 0.0031 \pm 0.0026 $&$                     $&$-1.7$\\
$R^{\bar\nu}$  &     CDHS       &$   0.384  \pm 0.016  \pm 0.007  $&$   0.3859 \pm 0.0003 $&$-0.1$\\
$R^{\bar\nu}$  &     CHARM      &$   0.403  \pm 0.014  \pm 0.007  $&$                     $&$ 1.1$\\
$R^{\bar\nu}$  &     CDHS 1979  &$   0.365  \pm 0.015  \pm 0.007  $&$   0.3813 \pm 0.0003 $&$-1.0$\\
\hline
$g_V^{\nu e}$  &     CHARM II   &$  -0.035  \pm 0.017             $&$  -0.0395 \pm 0.0004 $&  --- \\
$g_V^{\nu e}$  &      all       &$  -0.041  \pm 0.015             $&$                     $&$-0.1$\\
$g_A^{\nu e}$  &     CHARM II   &$  -0.503  \pm 0.017             $&$  -0.5063 \pm 0.0002 $&  --- \\
$g_A^{\nu e}$  &      all       &$  -0.507  \pm 0.014             $&$                     $&$-0.1$\\
\hline
$Q_W({\rm Cs})$&     Boulder    &$ -72.41   \pm 0.25\pm 0.80      $&$ -73.10   \pm 0.04   $&$ 0.8$\\
$Q_W({\rm Tl})$&Oxford + Seattle&$-114.8    \pm 1.2 \pm 3.4       $&$-116.7    \pm 0.1    $&$ 0.5$\\
\hline
\end{tabular}
\end{center}
\end{table}

Last year the Boulder group~\cite{Wood97} reported a much improved measurement
of the amplitude of the parity violating transition between Cesium's $6S$ and 
$7S$ states. The experimental error in the extracted weak charge, 
\be
  Q_W = - 72.41 \pm 0.25 \mbox{ (exp.)} \pm 0.80 \mbox{ (theory)}, 
\ee
decreased by a factor of 7. A slight improvement in the atomic theory 
calculations due to a new semi-empirical determination of the atomic 
polarizability~\cite{Dzuba97} shifted the central value of $Q_W$ closer to the
SM prediction and reduced the theory error. In the future the total error is 
expected~\cite{Wieman98} to decrease even further to about 0.5\%. $Q_W$ is 
obtained from an average of two different hyperfine transitions. The Boulder 
group observes a significant difference in the $6S_{F=4}$ to $7S_{F=3}$ and 
the $6S_{F=3}$ to $7S_{F=4}$ transitions. Most of this effect ($\approx 85\%$)
is assigned to the nuclear anapole moment (the axial electromagnetic form 
factor of the nucleus), which has not been observed before. Its size is 
somewhat larger than expected from theory estimates~\cite{Flambaum84}, but the 
latter are nuclear model dependent. Measurements of atomic parity violation 
(APV) in thallium~\cite{Edwards95} have also recently improved to match the 
theory calculations~\cite{Dzuba87} and to give useful constraints on new 
physics. Indeed, this type of measurement is very sensitive to the $S$ 
parameter to be discussed later.

The non-$Z$ pole precision observables are summarized in Table~\ref{nonzpole}.
They include the results from $\nu$-electron scattering by the 
CHARM II \col~\cite{Vilain94}, as well as the world averages. They are
presented in terms of the vector and axial-vector couplings,
$g_V^{\nu e}$ and $g_A^{\nu e}$. The $p\bar{p}$ value for $M_W$ assumes
a common systematic error of $\pm 50$~MeV between the three experiments.

\subsection{Other SM Tests}
Another important neutral current process in the context of new physics, and in
particular of low energy supersymmetry, is the flavor changing loop mediated 
transition $b \rightarrow s \gamma$. The ALEPH \col~\cite{Barate98} studies 
$B$ mesons and baryons, while CLEO~\cite{Alam95} focuses on meson decays. 
Many of the hadronic and systematic uncertainties cancel when one normalizes 
the rate by the charged current process $b \rightarrow c e \nu$. After having 
done so, one finds the two experiments in good agreement with each other, 
and the combined result,
\be
  R^{\rm exp} 
  = {\Gamma (b \rightarrow s \gamma)\over \Gamma (b \rightarrow c e \nu)}
  = (3.00 \pm 0.47) \times 10^{-3}.
\ee
A lot of effort has gone into theory calculations of $b \rightarrow s \gamma$.
For references and a discussion of electroweak radiative corrections 
see Ref.~\cite{Czarnecki98}, which quotes,
\be
  R^{\rm theory} = (3.01 \pm 0.25) \times 10^{-3},
\ee
for the SM. Alternatively, by writing
\be
  R^{\rm theory} = (3.00 \pm 0.25) (1 + 0.10 \rho) \times 10^{-3},
\ee
$R$ can also be interpreted as a measurement of one of the 
Wolfenstein parameters~\cite{Wolfenstein83} of the CKM mixing matrix,
\be
  \rho = 0.0 \pm 1.8.
\ee
The present world average~\cite{Caso98} of the muons anomalous magnetic moment 
is
\be
  a_\mu^{\rm exp} = {g_\mu - 2 \over 2} = (116592300 \pm 840) \times 10^{-11}.
\ee
On the other hand, the estimated SM electroweak 
contribution~\cite{Czarnecki96},
\be
  a_\mu^{\rm EW} = (151 \pm 4) \times 10^{-11},
\ee
is much smaller than the uncertainty. However, a new experiment at BNL 
is expected to reduce the experimental error to $\pm 40 \times 10^{-11}$
or better. The limiting factor will then be the uncertainty from the hadronic 
contribution~\cite{Davier98}, 
\be
  a_\mu^{\rm had} = (6924 \pm 62) \times 10^{-11},
\ee
which has recently been estimated with the help of $\tau$ decay data and 
finite-energy QCD sum rule techniques. This result constitutes a major 
improvement over previous ones which had more than twice the 
uncertainty~\cite{Eidelman95}. It would be important to verify it, and reduce 
the error even further to meet the experimental precision. 

Additional hadronic uncertainties are induced by the light-by-light scattering
contribution~\cite{Bijnens96},
\be
  a_\mu^{\rm LBLS} = ( - 92 \pm 32) \times 10^{-11},
\ee
and other subleading hadronic contributions~\cite{Krause97},
\be
  a_\mu^{\rm had} \left[\left({\alpha\over \pi}\right)^3\right] 
  = ( - 100 \pm 6) \times 10^{-11}.
\ee
The SM prediction is
\be
  a_\mu^{\rm theory} = (116591596 \pm 67) \times 10^{-11}.
\ee
With the anticipated accuracy at the BNL it will be possible 
to explore new physics up to energies of 5 TeV and more.
The result of the initial run at BNL in 1997 is~\cite{Karlen98}
\be
  a_\mu^{\rm BNL 97} = (1165925 \pm 15) \times 10^{-9}.
\ee
The error is expected to decrease to $\pm 1 \times 10^{-9}$ during the 
1998--1999 run. A precise measurement can give important hints and constraints
on new physics, specifically supersymmetry in the large $\tan\beta$ 
region~\cite{Carena97}.

\section{Theoretical Developments}
\label{theory}
\subsection{Schemes}
\label{schemes}
There are a variety of different renormalization schemes, reflected by various 
definitions of the weak mixing angle. The on-shell scheme is conceptually 
simple, carrying the tree level relation, $s^2_W \equiv 1 - M_W^2/M^2_Z$, 
to all orders. Due to large loop effects induced by the top-bottom doublet, 
$s^2_W$ is numerically several per cent smaller than the effective mixing 
angles, $\bar{s}^2_f \equiv \kappa_f s^2_W$. These are defined through $Z$ pole
asymmetries, and therefore (like $s^2_W$) (pseudo)-observable and scheme 
independent. The relatively large form factors $\kappa_f$ induce large 
reducible higher order corrections, which frequently dominate the irreducible 
(genuine) corrections. Numerically much closer to the $\bar{s}^2_f$ is the 
$\ms$ definition $\sin^2 \hat\theta_\ms (M_Z) \equiv \hat{s}^2_Z$. 
As a consequence the $\ms$ scheme has excellent convergence properties. 
It is also (unlike the $\bar{s}^2_f$) flavor independent, and it is very 
convenient in the context of gauge coupling unification. There it is compared 
to the strong coupling constant, $\alpha_s$, which is traditionally treated 
in the $\ms$ scheme, as well. In general, calculations in the $\ms$ scheme tend
to be technically simpler, and it is especially convenient in the presence of 
mixed QCD-electroweak corrections. One drawback of $\hat{s}^2_Z$ is that it is
a theoretical construct and not directly related to an observable. As a result,
there exist various versions of $\ms$ mixing angles. Here we use the definition
with the top quark decoupled~\cite{Marciano90}. The scale, which has to be 
specified as well, can sometimes be used to eliminate spurious logarithms 
in coefficient functions, a freedom not available for the other definitions.

\subsection{Recent Radiative Corrections}
\label{radcorr}
To match the accuracy of the high precision data, multi-loop perturbative 
calculations have to be performed. These include leading two-loop electroweak,
three-loop mixed electroweak-QCD, and three-loop QCD corrections. 
${\cal O} (\alpha \alpha_s)$ vertex corrections to $Z$ 
decays~\cite{Czarnecki96A} have become available only recently, inducing 
an increase in the extracted $\alpha_s$ by about 0.001. The inclusion of top 
mass enhanced two-loop ${\cal O} (\alpha^2 m_t^4)$~\cite{Barbieri92} and 
${\cal O} (\alpha^2 m_t^2)$~\cite{Degrassi96} effects is crucial for a reliable
extraction of $M_H$.

We have collected all available results in a new radiative correction
package. All $Z$ pole and low energy observables are selfconsistently
evaluated with common inputs. The routines are written entirely within 
the $\ms$ scheme, using $\ms$ definitions for all gauge couplings and 
quark masses. This reduces the size of higher order terms in the QCD expansion.

The largest theoretical uncertainty arises from the $M_W$--$M_Z$--$\hat{s}^2_Z$
interdependence. The problem is directly related to the renormalization group 
(RGE) running of the electromagnetic coupling,
\be
   \alpha (M_Z) = {\alpha\over 1 - \Delta\alpha (M_Z)}.
\ee
While the contributions from leptons and bosons (and the top quark when not 
technically decoupled) can be computed with sufficient accuracy, the hadronic 
contributions from the five lighter quarks escape a first principle treatment 
due to strong interaction effects. If one trusts perturbative QCD down to 
energies below the $J/\Psi$ threshold, one can use RGE techniques for charm and
bottom quarks, as well~\cite{Erler98A}. Complications from quark confinement 
are then effectively moved to the actual extraction of the $c$ and $b$ quark 
masses, which contribute a major uncertainty in this approach. Alternatively, 
one can use~\cite{Davier98,Zeppenfeld95} $J/\Psi$ and $\Upsilon$ resonance 
parametrizations and a dispersion relation (DR) to arrive at the charm and 
bottom contributions to $\Delta\alpha (M_Z)$. The latter approach is slightly 
more precise, while the first one allows a selfconsistent treatment of the 
parametric uncertainty from $\alpha_s$, an advantage in electroweak fits. 
Both approaches give very similar results. Finally, one might prefer to use 
$e^+ e^-$ data and the dispersive approach for the entire energy regime up to 
well above the $\Upsilon$ resonances (the cutoff is typically 
40 GeV)~\cite{Eidelman95,Burkhardt95,Alemany98}. However, since the relevant
data is not very precise, one faces larger uncertainties; the results are in 
good agreement with the more theoretical approaches, but tend to give larger
values of $\Delta\alpha (M_Z)$ and smaller resulting best fit values for $M_H$.
As for the three light flavors, a DR cannot be circumvented in any of the 
approaches, and differences arise only due to different data or fit procedures.
The calculation of $\hat\alpha_\ms$ is naturally performed using an 
unsubtracted DR~\cite{Erler98A}; the on-shell coupling is computed through a 
subtracted one~\cite{Davier98,Eidelman95,Zeppenfeld95,Burkhardt95,Alemany98}. 
In our analysis for the Particle Data Group~\cite{Erler98B} we relied on 
Ref.~\cite{Alemany98}, while here we use~Ref.~\cite{Erler98A}. The LEP 
Electroweak Working Group (LEPEWWG)~\cite{Karlen98} employs 
Ref.~\cite{Eidelman95}. 

It is interesting that the TOPAZ collaboration at TRISTAN~\cite{Levine97} has
actually observed the running of $\alpha$, albeit with large uncertainty. 
From the cross section for $e^+ e^- \rightarrow e^+ e^- \mu^+ \mu^-$ 
(relative to $\mu^+ \mu^-$) they obtain $\alpha^{-1} (57.77 \mbox{ GeV}) 
= 128.5 \pm 1.9$, compared with the theoretical expectation $129.6 \pm 0.1$.

\section{Standard Model Fit Results}
\label{sm}
\subsection{Overview}
We use the complete data set described in Section~\ref{newdata}, and 
summarized in Tables~\ref{zpole} and~\ref{nonzpole} for a global electroweak 
analysis. We carefully took into account experimental and theoretical
correlations, in particular in the $Z$-lineshape sector, the heavy flavor 
sector from LEP and the SLC, and for the DIS experiments. 
Predictions within and beyond the SM were calculated by means of a new 
radiative correction program based on the $\ms$ renormalization scheme 
(see Section~\ref{theory}). All input and fit parameters are included in a 
selfconsistent way, and the correlation (present in theory evaluations of 
$\alpha (M_Z)$) between $\alpha_s$ and $\Delta\alpha (M_Z)$ is automatically
taken care of. We find very good agreement with the results of the 
LEPEWWG~\cite{Karlen98}, except for well-understood effects originating from
higher orders. We would like to stress that this agreement is quite remarkable
as they use the electroweak library ZFITTER~\cite{Bardin92}, which is based on 
the on-shell renormalization scheme. It also demonstrates that once the most 
recent theoretical calculations, in particular 
Refs.~\cite{Czarnecki96A,Degrassi96}, are taken into account, the theoretical 
uncertainty becomes quite small, and is in fact presently negligible compared 
to the experimental errors. The relatively large theoretical uncertainties 
obtained in the Electroweak Working Group Report~\cite{Bardin97} were 
estimated using different electroweak libraries, which did not include the 
full range of higher order contributions available now. 

\begin{table}
\caption{Fit values of $\hat{s}\,^2_Z$, $s_W^2$, $\alpha_s$, and $m_t$ for 
various combinations of observables. The quoted values are for the global 
best fit result, $M_H = 107~$ GeV. The uncertainties are from $m_t$, 
$\alpha_s$, and $\alpha (M_Z)$.}
\label{s2z}
\vspace{0.2cm}
\begin{center}
\footnotesize
\begin{tabular}{|l|l|l|l|l|}
\hline Data & $\hat{s}^2_Z$ & $s^2_W$ & $\alpha_s (M_Z)$ & $m_t$ [GeV] \\ 
\hline
All indirect + $m_t$          & 0.23129(12) & 0.2233(4)  & 0.1206(30)           & $171.4 \pm 3.4$ \\
All indirect + $m_t+\alpha_s$ & 0.23126(12) & 0.2233(4)  & 0.1188(18)           & $171.4 \pm 3.4$ \\
All indirect                  & 0.23134(15) & 0.2236(5)  & 0.1206(30)           & $169.4 \pm 4.6$ \\
$Z$ pole                      & 0.23136(17) & 0.2236(6)  & 0.1206(30)           & $168.6 \pm 5.3$ \\
LEP 1                         & 0.23163(21) & 0.2246(7)  & 0.1219(31)           & $161.2 \pm 6.4$ \\
SLD + $M_Z$                   & 0.23080(31) & 0.2216(11) & 0.1200 ({\rm fixed}) & $185.8 \pm 9.0$ \\
$A_{FB} (b)$ + $M_Z$          & 0.23198(39) & 0.2259(15) & 0.1200 ({\rm fixed}) & $148.5 \pm 13.5$\\
$M_W$ + $M_Z$                 & 0.23115(34) & 0.2228(12) & 0.1200 ({\rm fixed}) & $175.4 \pm 10.0$\\
DIS + $M_Z$                   & 0.23187(78) & 0.2255(29) & 0.1200 ({\rm fixed}) & $152.2 \pm 25.1$\\
DIS + $m_t$                   & 0.2334(22)  & 0.2253(21) & 0.1200 ({\rm fixed}) & $173.9 \pm 5.0$ \\
\hline
\end{tabular}
\end{center}
\end{table}

In the Standard Model analysis we use the fine structure constant, $\alpha$,
and the Fermi constant, $G_F =  1.16637 (1) \times 10^{-5}$ GeV$^{-2}$, as 
fixed inputs. The error in $G_F$ is now of purely experimental origin after 
the very recent calculation of the two-loop QED corrections to $\mu$ decay have
been completed~\cite{vanRitbergen98}. They lower the central value by 
$2\times 10^{-10}$ GeV$^{-2}$ and the extracted $M_H$ by 1.3\%. Moreover, there
are five independent fit parameters, which can be chosen to be $M_Z$, $M_H$, 
$m_t$, $\alpha_s$, and $\Delta\alpha (M_Z)$. Alternatively, $M_Z$ can be 
replaced by $s^2_W$ or $\hat{s}^2_Z$. Unless explicitly stated, we do not use 
$\alpha_s$ determinations from outside the $Z$ lineshape sector. The fit to all
precision data is perfect with an overall $\chi^2 = 28.8$ for 36 degrees of 
freedom, and yields,
\be
\ba{lcc}
\label{fitresults}
           M_H &=& 107^{+67}_{-45} \mbox{ GeV}, \\
           m_t &=& 171.4  \pm 4.8  \mbox{ GeV}, \\
      \alpha_s &=& 0.1206 \pm 0.0030, \\
   \hat{s}^2_Z &=& 0.23129 \pm 0.00019, \\
\bar{s}^2_\ell &=& 0.23158 \pm 0.00019, \\
         s^2_W &=& 0.22332 \pm 0.00045.
\ea
\ee
Moreover, none of the observables deviates from the SM best 
fit prediction by more than 2 standard deviations. 

In Table~\ref{s2z} we show the results of fits to various data sets. 
Shown are the $\ms$ and on-shell mixing angles, as well as $\alpha_s$ 
and $m_t$. In these fits we have fixed the Higgs mass to its global best 
fit value, $M_H = 107$~GeV, so that the sensitivity to $m_t$ becomes 
transparent. The first line corresponds to the fit to all data. Note the 
smaller errors compared to the results~(\ref{fitresults}) where $M_H$ was 
a free parameter. The third line corresponds to a fit to all indirect data. 
The extracted $m_t = 169.4 \pm 4.6$~GeV is in good agreement with the 
direct result from the Tevatron, and is of similar accuracy. On the other 
hand, the $m_t$ extracted from the LEP~1 (SLD) observables is slightly 
below (above) the Tevatron value. There is still a $2.2 \sigma$ discrepancy 
between the $\hat{s}^2_Z$ determinations at LEP and the SLC. The last 2 
lines show the results from DIS. When combined with $M_Z$, DIS represents 
a measurement of $\hat{s}^2_Z$ ($R_\nu$), and also has some sensitivity 
to $m_t$ ($R_{\bar\nu}$). The last fit is equivalent to regarding DIS as a 
measurement of $s^2_W$. Even after the end of the LEP~1 era, $\nu$-hadron 
scattering experiments continue to represent competitive measurements. 
This is even more true in the presence of new physics. 

\subsection{$M_H$}
\label{mh}
The data show a strong preference for a low $M_H \sim {\cal O} (M_Z)$. 
Unlike in previous analyses, the central value of the global fit 
to all precision data, including $m_t$ and excluding further 
constraints from direct searches,
\be
\label{mh_fit}
   M_H = 107^{+67}_{-45} \mbox{ GeV},
\ee
is now above the direct lower limit,
\be
\label{mh_limit}
   M_H > 90 \mbox{ GeV [95\% CL]},
\ee
from searches at LEP 2~\cite{McNamara98}. In fact, it coincides with the
$5 \sigma$ discovery limit (and is 2 GeV below the 95\% exclusion limit)
from LEP 2 running at 200~GeV center of mass energy with 200 pb$^{-1}$
integrated luminosity per experiment~\cite{McNamara98}. The 90\% central 
confidence interval from precision data only is given by
\be
  39 \mbox{ GeV} < M_H < 226 \mbox{ GeV}.
\ee

These results are to be compared with the theoretical SM expectation,
\be
   115 \mbox{ GeV} \lsim M_H \lsim 750 \mbox{ GeV},
\ee
where the lower bound is derived from vacuum stability 
requirements~\cite{Sher93}, and the upper bound is a (lattice) triviality 
bound~\cite{Heller93}. The fit result~(\ref{mh_fit}) is also in agreement 
with the prediction for the lightest neutral Higgs boson~\cite{Heinemeyer98},
\be
\label{mh0}
   m_{h^0} \lsim 130 \mbox{ GeV},
\ee
within the Minimal Supersymmetric Standard Model (MSSM). In non-minimal
extensions of the MSSM, the bound~(\ref{mh0}) is relaxed to about 150 GeV.

For the determination of the proper $M_H$ upper limits, we scan equidistantly 
over $\ln M_H$, combining the likelihood $\chi^2$ function from the 
precision data with the exclusion curve (interpreted as a prior probability 
distribution function) from LEP 2~\cite{McNamara98}. This curve is from Higgs 
searches at center of mass energies up to 183 GeV. We find the 90 (95, 99)\% 
confidence upper limits,
\be
  M_H < 220 \mbox{ (255, 335) GeV}.
\ee
Notice, that the LEP 2 exclusion curve increases the 95\% upper limit by
almost 30 GeV. 

Indirect $M_H$ constraints from precision data are now very similar in 
precision to the indirect $m_t$ constraints from less than a decade 
ago~\cite{Langacker89}, just before the commencement of the $Z$ pole era at 
LEP~1 and the SLC. This is rather remarkable, as $M_H$ effects are, in contrast
to the leading quadratic $m_t$ effects, only approximately logarithmic. Higgs 
hunting in precision data is further hampered by the strong correlation of 
$\ln M_H$ with $m_t$ (70\%) and $\alpha (M_Z)$ (30\%). The strongest 
constraints come from the asymmetries, but $M_W$, $R_b$, and the lineshape 
observables are also significant. In the past, the tendency for a low $M_H$ 
came almost entirely~\cite{Erler95A} from $A_{LR}$ and $R_b$, both of which 
were in conflict with the SM. Now they deviate much less (especially $R_b$), 
plus there are extra constraints. This also reduces possible confusion with
new physics (to which $R_b$ is particularly sensitive). It should be noted,
however, that the results on $M_H$ are strongly correlated with the $S$ and
$T$ parameters, as discussed in Section~\ref{np}. The increase in $\chi^2$ 
when shifting $M_H$ up to 1~TeV, which used to be only a few units, is now
\be
   \Delta \chi^2 = \chi^2 (M_H = 1 \mbox{ TeV}) - \chi^2_{\rm min} = 25.7,
\ee
i.e., a TeV scale SM Higgs boson is now excluded at the $5 \sigma$ level. 

On the other hand, the $\chi^2$ function is fairly shallow for Higgs masses 
up to ${\cal O} (200 \mbox{ GeV})$. Precise predictions are difficult due 
to the low sensitivity in that regime. One also has to keep in mind that there
are still some $2 \sigma$ level deviations, such as in $A_{LR}$ and 
$A_{FB} (b)$, and ambiguities in the treatment of $\alpha(M_Z)$. Unlike the
central values, however, the upper limits on $M_H$ are rather insensitive 
to the used $\alpha(M_Z)$~\cite{Erler98}. This is due to compensating
effects from the larger central value of $\alpha (M_Z)$ (corresponding to 
lower extracted Higgs masses) and the larger error bars in the data
driven approach~\cite{Eidelman95,Burkhardt95,Alemany98} (see the
discussion in Section~\ref{radcorr}).

One can take the point of view~\cite{Chanowitz98} that the $A_{LR}$ measurement
at the SLC and $A_{FB} (\tau)$ from LEP are by themselves in conflict with the 
lower Higgs limit~(\ref{mh_limit}). For example, using the Tevatron $m_t$ 
one would predict from $A_{LR}$,
\be
   M_H (A_{LR}) = 39^{+46}_{-25} \mbox{ GeV}.
\ee
This may well be due to a statistical fluctuation, and to avoid a bias we 
treat it as such. Indeed, $A_{LR}$ and $A_{FB} (\tau)$, are both consistent 
with the lower direct $M_H$ limit within $1.1\sigma$. Moreover the $\chi^2$ 
per d.o.f.\ in the SM fit is excellent, discouraging the use of scale factors 
to increase error bars. It should also be stressed again that our upper $M_H$ 
limits do take into account the direct search results in a proper Bayesian way.
However, to get a sense of how sensitive our limits depend on $A_{LR}$ and 
$A_{FB} (\tau)$, we study the effect of repudiating {\em both\/} measurements 
completely from the fit, and find,
\be
   M_H (\hspace{2pt}\not\hspace{-5pt}A_{LR}, \hspace{5pt}\not\hspace{-5pt}
   A_{FB}(\tau)) = 195^{+119}_{-78} \mbox{ GeV},
\ee
and the 90 (95)\% upper limit,
\be
  M_H (\hspace{2pt}\not\hspace{-5pt}A_{LR}, \hspace{5pt}\not\hspace{-5pt}
   A_{FB}(\tau)) < 356 \mbox{ (419) GeV}.
\ee
Hence, even such a radical treatment increases the upper $M_H$ limit by only
about 60\% or $\sim 150$~GeV. 

We generally disagree with the conclusions drawn from PDG scale factor 
studies~\cite{Chanowitz98} in which 95\% upper limit of ${\cal O} (700)$~GeV 
are found. These analyses were based on data presented at summer~97 and 
spring~98 conferences, and should therefore be compared with our earlier 
results published in Refs.~\cite{Erler98,Erler98B} with upper limits of 
${\cal O} (300)$~GeV. Since then (see the footnote in Section~\ref{zpolephys})
there are new results on $A_{LR}$, and also $A_{FB}(\tau)$ has somewhat 
changed. When using the summer~98 conferences the author of 
Refs.~\cite{Chanowitz98} agrees with our results~\cite{Chanowitz98A}.

\subsection{$\alpha_s (M_Z)$}
\label{alphas}
As for the extraction of the strong coupling constant at the $Z$ scale,
we find the best fit value\footnote{For comparison,
the LEPEWWG~\cite{Karlen98} quotes $\alpha_s = 0.119 \pm 0.003$.},
\be
\label{alpha_lineshape}
   \alpha_s = 0.1206 \pm 0.0030,
\ee
in excellent agreement with other determinations. For example, the 
ALEPH~\cite{Barate98A} and OPAL~\cite{Ackerstaff98A}~\col s obtain from 
hadronic $\tau$ decays,
\be
\ba{c}
   \alpha_s (\Gamma_\tau) = 0.1202 \pm 0.0027, \\
   \alpha_s (\Gamma_\tau) = 0.1219 \pm 0.0020,
\ea
\ee
respectively. Another result of similar precision is obtained from $\Upsilon$ 
spectroscopy using non-relativistic QCD for lattice gauge 
theory~\cite{Davies97}, which is also consistent with a preliminary result 
from $J/\Psi$ spectra~\cite{Shigemitsu97},
\be
\ba{r}
   \alpha_s (b\bar{b} \mbox{ spectrum}) = 0.1174 \pm 0.0024, \\
   \alpha_s (c\bar{c} \mbox{ spectrum}) = 0.1159 \pm 0.0030.
\ea
\ee
Measurements of the proton structure functions $F_2$ and $x F_3$ in neutrino 
DIS yield~\cite{Seligman97}
\be
   \alpha_s ({\rm DIS}) = 0.119 \pm 0.002 \pm 0.004,
\ee
where the first error is experimental and the second theoretical. 
The NuTeV \col\ plans to reduce the total error of $\alpha_s$ from scaling 
violations in DIS to 0.002. Finally, there are a variety of jet event shape 
extractions of $\alpha_s$ from $e^+ e^-$ annihilation at, below, and above 
the $Z$ peak, again in excellent agreement with Eq.~(\ref{alpha_lineshape}). 
One can clearly conclude that the old low energy versus high energy 
controversy for $\alpha_s$ is over! 

We can use these $\alpha_s$ measurements as an additional external constraint 
in global fits. In order to do so, we use the world average by the Particle
Data Group~\cite{Hinchliffe98} with the $Z$ lineshape value removed,
\be
\label{alphas_hinchliffe}
  \alpha_s = 0.1178 \pm 0.0023.
\ee
The result (cf., the second row in Table~\ref{s2z}),
\be
  \alpha_s = 0.1188 \pm 0.0018,
\ee
can be viewed as the present world average. Inclusion of
the constraint~(\ref{alphas_hinchliffe}) reduces the 90 (95, 99)\% upper
$M_H$ limit by 6 (9, 13) GeV. 

The lineshape value~(\ref{alpha_lineshape}) is also consistent with predictions
from gauge unification in supersymmetric GUT's and string theories. 
Using $\alpha (M_Z)$ and $\hat{s}_Z^2$ one predicts~\cite{Langacker95A}
\be
   \alpha_s = 0.130 \pm 0.010.
\ee

\subsection{Future Prospects}
\label{fp}
The LEP~1 results are now close to being finalized, although some work on
systematic errors and correlations still needs to be done. Similarly, the
uncertainty in $M_W$ from the Tevatron run~I is expected to decrease somewhat. 
Run~II at the Tevatron with its ten times larger luminosity is anticipated 
to measure $M_W$ to $\pm 40$~MeV per experiment and channel. The statistical 
error of $M_W$ from LEP~2 is likely to decrease to $\pm 40$~MeV within a year 
and to $\pm 25$~MeV after its completion. For a projection of the systematic 
error, the effects of color reconnection and Bose-Einstein correlations need 
to be understood more rigorously. Future $M_W$ measurements could reach 
a precision of $\pm 25$~MeV, including a theoretical uncertainty from 
uncalculated higher order radiative corrections. Moreover, run~II should 
determine $m_t$ within 2~GeV, including the theoretical ambiguity from the 
conversion between pole and running mass definitions. The SLD \col\ is seeking
for additional run time (``SLD 2000''), which would permit them to double their
statistics. This would greatly improve the $A_{LR}$ and $A_{LR}^{FB}$ 
measurements, which are still statistically limited. Given the superior 
three-dimensional vertexing with the SLD detector (VXD3), it would also allow 
competitive measurements of $R_b$ and $R_c$, with errors comparable to those 
from LEP. 

Adding the hypothetical constraints,
\be
\ba{llrr}
  m_t    &\mbox{ (run II) }         &=& 171.4    \pm 2.0   \mbox{ GeV}, \\
  M_W    &\mbox{ (run II + LEP 2) } &=&  80.362  \pm 0.025 \mbox{ GeV}, \\
  A_\ell &\mbox{ (SLD 2000) }       &=&   0.1466 \pm 0.0020, \\
  R_b    &\mbox{ (SLD 2000) }       &=&   0.2158 \pm 0.0010, \\
\ea
\ee
to the data (where the central values are the current global best fit values),
one might find around the year 2002,
\be
   M_H \mbox{ (future)} = 107^{+36}_{-29} \mbox{ GeV},
\ee
i.e., a 30\% determination. From direct searches, Higgs masses up to 
94 (97)~GeV can be discovered (excluded) from the present 189~GeV run at LEP~2.

After run~II there might be a further luminosity upgrade at the Tevatron
(TeV33) reducing the top mass error to about $\pm 1$~GeV. High precision
measurements of the total width and the leptonic branching ratio of the $W$ 
boson would be possible. In addition, a per mille determination of the weak 
mixing angle through $A_{FB} (\ell)$ is conceivable. Most importantly, with
30~fb$^{-1}$ at TeV33, Higgs boson searches up to 130~GeV would be possible.

\section{Beyond the Standard Model}
\label{np}
\subsection{Unification or Compositeness}
The successful supersymmetric gauge coupling unification discussed in 
Section~\ref{alphas} could be coincidental.  If taken seriously, however,
it could also be taken as circumstantial evidence for supersymmetry (SUSY).
Moreover, it would constrain extra matter to come either in complete standard 
families or singlets.  And it would demand the absence of extended gauge 
structures below the unification scale $M_{GUT} \sim 10^{16}$ GeV, unless they
commute with the SM gauge group. Of course, there is always the possibility of
subtle cancellations of different effects\footnote{For a recent example, see
Ref.~\cite{Cleaver98}.}.

In addition to the encouraging gauge unification, there is the agreement
between the general prediction from perturbative low energy SUSY,
\be
\label{susy_mh}
   m_{h^0} \lsim 150 \mbox{ GeV},
\ee
and the precision data. The results in Section~\ref{mh} apply strictly only
in the context of the SM, or when the extra sparticles and Higgs bosons
predicted by SUSY are decoupled, i.e., heavier than a few hundred GeV.
In this case, effects in the precision data are small, and it is a remarkable
prediction of the decoupled MSSM that no deviation in the precision data are
to be expected. Similarly, flavor changing neutral currents (FCNC) and CP 
violating effects may be small, as well. 

On the contrary, in scenarios of new physics involving a composite (dynamical)
Higgs sector and/or composite fermions, one expects a variety of effects. These
include in particular FCNC, which are typically predicted along with many
new 4-Fermi operators. However, (in the absence of fine tuning) FCNC operators
are already excluded up to scales of ${\cal O} (100 \mbox{ TeV})$, and APV 
excludes contact operators of $eq$ type for scales up to 
${\cal O} (10 \mbox{ TeV})$. These types of new physics also tend to predict 
a decrease in $R_b$, the opposite of what is being observed. Finally, they are 
in conflict with oblique radiative corrections, as defined and discussed below.

\subsection{Oblique Parameters}
The data are precise enough to constrain additional parameters describing 
physics beyond the SM. Of particular interest is the $\rho$-parameter, defined 
by
\be
  \rho_0 = {M_W^2 \over M_Z^2 \hat{c}^2_Z \hat{\rho} (m_t,M_H)},
\ee
which is a measure of the neutral to charged current interaction strength. 
The SM contributions are absorbed in $\hat{\rho}$, so that in the SM 
$\rho_0 = 1$, by definition. Examples for sources of $\hat{\rho} \neq 1$ 
include non-degenerate extra fermion or boson doublets, and non-standard Higgs
representations. 

In a fit to all data with $\rho_0$ as an extra fit parameter, the correlation
of $M_H$ with $m_t$ is lifted, and replaced by a strong (73\%) 
correlation\footnote{$\rho_0$ is also strongly anticorrelated with $\alpha_s$ 
($-53\%$) and $m_t$ ($-46\%$).} with $\rho_0$. As a result upper limits on 
$M_H$ are weak when $\rho_0$ is allowed. Indeed, $\chi^2 (M_H)$ is very 
shallow with
\be
   \Delta \chi^2 = \chi^2 (1 \mbox{ TeV}) - \chi^2 (M_Z) = 4.5,
\ee
and its minimum is at $M_H = 46$~GeV, which is already excluded. We obtain,
\be
\ba{lcr}
\label{rhofit}
         \rho_0 &=& 0.9996^{+0.0009}_{-0.0006}, \vspace{2pt} \\
            m_t &=& 172.9   \pm 4.8 \mbox{ GeV}, \\
       \alpha_s &=& 0.1212  \pm 0.0031,
\ea
\ee
in excellent agreement with the SM. The central values are for $M_H = M_Z$, 
and the uncertainties are $1 \sigma$ errors and include the range,
$M_Z \leq M_H \leq 167$~GeV, in which the minimum $\chi^2$ varies within one 
unit. Note, that the uncertainties for $\ln M_H$ and $\rho_0$ are non-Gaussian:
at the $2 \sigma$ level ($\Delta \chi^2 \leq 4$), Higgs masses up to 800~GeV 
are allowed, and we find
\be
   \rho_0 = 0.9996^{+0.0031}_{-0.0013} \mbox{ ($2 \sigma$)}.
\ee
This implies strong constraints on the mass splittings of extra fermion and 
boson doublets~\cite{Veltman77}, 
\be
  \Delta m^2 = m_1^2 + m_2^2 - \frac{4 m_1^2 m_2^2}{m_1^2 - m_2^2} 
               \ln {m_1\over m_2} \geq (m_1 - m_2)^2,
\ee
namely, at the $1\sigma$ and $2\sigma$ levels, respectively,
\be
\label{splittings}
   \sum\limits_i {C_i\over 3} \Delta m^2_i < \mbox{ (38 GeV)}^2 
   \mbox{ and (93 GeV)}^2,
\ee
where $C_i$ is the color factor. Due to the general condition~(\ref{susy_mh}) 
in the MSSM, stronger $2 \sigma$ constraints result here,
\be
   \rho_0 \mbox{ (MSSM) } = 0.9996^{+0.0017}_{-0.0013} \mbox{ ($2 \sigma$)}.
\ee
The constraints~(\ref{splittings}) would therefore change to
\be
   \sum\limits_i {C_i\over 3} \Delta m^2_i < \mbox{ (38 GeV)}^2 
   \mbox{ and (64 GeV)}^2 \mbox{ (MSSM)}.
\ee

Similarly, constraints on heavy degenerate chiral fermions can be obtained 
through the $S$ parameter~\cite{Peskin90}, defined through a difference of 
$Z$ boson self-energies,
\be
  \frac{\hat\alpha (M_Z) }{4 \hat{s}_Z^2 \hat{c}_Z^2} S \equiv 
  \frac{\Pi^{\rm new}_{ZZ} (M_Z^2) - \Pi^{\rm new}_{ZZ} (0) }{M_Z^2}.
\ee
The superscripts indicate that $S$ includes new physics contributions only.
Likewise, $T = (1 - \rho_0^{-1})/\hat\alpha$ and the $U$ parameter to be
discussed below, also vanish in the SM\footnote{Thus, our definition differs
somewhat from the original definition~\cite{Peskin90} which included the
$m_t$ and $M_H$ contributions to the self-energies in $S$, $T$, and $U$.}. 
A fit to all data with $S$ allowed yields,
\be
\ba{lcr}
\label{Sfit}
              S &=& -0.20^{+0.24}_{-0.17}, \vspace{4pt} \\
            M_H &=& 390^{+690}_{-310} \mbox{ GeV}, \vspace{2pt} \\
            m_t &=& 172.9   \pm 4.8 \mbox{ GeV}, \\
       \alpha_s &=& 0.1221  \pm 0.0035.
\ea
\ee
In the presence of $S$, constraints on $M_H$ virtually disappear. In fact,
$S$ and $M_H$ are almost perfectly anticorrelated ($-92\%$). 
A heavy degenerate ordinary or mirror family contributes $2/3\pi$ to $S$.
By requiring $M_Z \leq M_H \leq 1$~TeV, we find with $3\sigma$ confidence,
\be
   S = -0.20^{+0.40}_{-0.33} \mbox{ ($3 \sigma$)}.
\ee
A fourth sequential fermion family is excluded at the 99.8\% CL.

New physics contributions to the third oblique parameter, $U$, which is
defined through
\be
   \frac{\hat\alpha(M_Z)}{4 \hat{s}_Z^2} \left( S + U \right) \equiv 
   \frac{\Pi^{\rm new}_{WW} (M_W^2) - \Pi^{\rm new}_{WW} (0) }{M_W^2},
\ee
are usually expected to be small. A fit to all data with $U$ allowed,
\be
\ba{lcr}
\label{Ufit}
              U &=& 0.09 \pm 0.19, \vspace{2pt} \\
            M_H &=& 110^{+70}_{-46} \mbox{ GeV}, \vspace{2pt} \\
            m_t &=& 171.1   \pm 4.9 \mbox{ GeV}, \\
       \alpha_s &=& 0.1207  \pm 0.0030,
\ea
\ee
reveals perfect agreement with the SM prediction $U = 0$. Notice, that 
allowing $U$ has little effect on the extracted $M_H$, as it has only 
small correlations with the SM parameters. 

A simultaneous fit to $S$, $T$, and $U$ can be performed only relative to a 
specified $M_H$. If one fixes $M_H = 600$~GeV, as is appropriate in QCD-like 
technicolor models, one finds
\be
\ba{rcr}
   S &=& -0.27 \pm 0.12, \\
   T &=&  0.00 \pm 0.15, \\
   U &=&  0.19 \pm 0.21.
\ea
\ee
Notice, that in such a fit the $S$ parameter is significantly smaller than 
zero. From this an isodoublet of technifermions, assuming $N_{TC} = 4$ 
technicolors, is excluded by almost 6 standard deviations, and a full 
technigeneration by more than $15\sigma$. However, the QCD-like models are 
excluded on other grounds, such as FCNC. In particular, in models of walking 
technicolor $S$ can be smaller or even negative~\cite{Gates91}. 

The allowed range of the oblique parameters in the context of SUSY
is obtained by demanding $M_Z \leq M_H \leq 150$~GeV, which yields,
\be
\ba{rcr}
   S &=& -0.17^{+0.17}_{-0.12}, \vspace{4pt} \\
   T &=& -0.16^{+0.15}_{-0.18}, \vspace{4pt} \\
   U &=&  0.19 \pm 0.21.
\ea
\ee
Note the $2\sigma$ upper limit $T \leq 0.14$. Allowing supersymmetric
contributions to $R_b$, which can be mediated by light top squark and
chargino loops, this limit would tighten further to 
\be
\label{T_limit}
   T \leq 0.12 \mbox{ ($2 \sigma$)}.
\ee
These results are to be compared with the predictions of various scenarios 
for the mediation of SUSY breaking from the hidden to the observable 
sector. For example, in the minimal supergravity model with universal soft 
SUSY breaking terms, there are regions of parameter space in which $T$ can be 
as large as 0.20, so they have to be excluded. Of course, there are in general
also (smaller) contributions to $S$ and $U$, as well as non-oblique 
corrections, so much more parameter space can be excluded than what is 
suggested by the constraint~(\ref{T_limit}). A systematic analysis of precision
data in the MSSM, and a discussion of the excluded parameter space can be found
in Ref.~\cite{Erler98C}.

\subsection{Weaker Interactions}
Many GUT's and string models predict extra gauge symmetries and new exotic 
states. For example, $SO(10)$ GUT contains an extra $U(1)$ as can be seen
from its maximal subgroup, $SU(5) \times U(1)_\chi$. The spinorial ${\bf 16}$
representation contains besides the known quarks and leptons one further state,
the right-handed neutrino. This is highly desirable for the generation of 
neutrino mass through the see-saw mechanism. However, the scale at which the 
$U(1)_\chi$ is broken is not predicted, and the mass of the accompanying gauge
boson, $M_{Z^\prime}$, is arbitrary. Thus, there is no 
reason to expect a $Z^\prime$ of this kind at the electroweak or TeV scales, 
and to look for them there is like the ``search under a lamppost''. 

Similarly, $E_6$ GUT contains the subgroup $SO(10) \times U(1)_\psi$, giving
rise to another $Z^\prime$. In addition, the fundamental {\bf 27} contains
besides the ${\bf 16}$ of $SO(10)$, a fundamental decouplet plus a singlet.
The decouplet decomposes into a vector-like pair of $SU(5)$ fundamentals,
containing an isoscalar down-type quark and a lepton doublet~\cite{Achiman78}.
Searches for these exotic states at LEP~2 and the Tevatron set lower limits 
of about 90~GeV on their masses. 

The potential $Z^\prime$ boson is in general a superposition of the SM $Z$ 
and the new boson associated with the extra $U(1)$,
\be
\ba{lcr}
   Z        &=&   Z_1^0 \cos\theta + Z_2^0 \sin\theta, \\
   Z^\prime &=& - Z_1^0 \sin\theta + Z_2^0 \cos\theta, 
\ea
\ee
where~\cite{Langacker84}
\be
   \tan^2\theta = \frac{M_{Z_1^0}^2 - M_Z}{M_{Z^\prime} - M_{Z_1^0}^2},
\ee
and where $M_{Z_1^0}$ is the SM value for $M_Z$ in the absence of mixing. 
Note that $M_Z < M_{Z_1^0}$ (for the case that $Z_1^0$ is the lighter of the
states before mixing), and that the SM $Z$ couplings are changed by the
mixing. $M_{Z_1^0}$ can be calculated from $M_W$ and compared with $M_Z$
from LEP. A significant difference could be an indication for the presence
of the extra boson, as would a difference between $M_Z$ and the value predicted
by the other $Z$ pole observables. $Z^\prime$ exchange is suppressed (out of 
phase) at the $Z$ pole. However, stringent limits can be obtained from weak 
neutral current data at lower energies. 

One finds $M_{Z_\chi} > 330$~GeV from precision data for arbitrary mixing, 
but limits can increase to the TeV scale in specific models with known
mixing. Collider limits on $Z^\prime$ masses depend on the chiral
couplings of the new gauge boson to ordinary quarks and leptons, and on
possible exotic decay modes.  For a $Z^\prime$ with SM couplings and 
a width which scales like the SM one, CDF (D\O) sets a lower 
limit~\cite{Dorigo98} of 690 (670)~GeV. For typical GUT models the limits
are $M_{Z_\chi} > 600$--1000~GeV, with very constrained mixing angles $\theta$
of at most a few per mille. Limits on a leptophobic $Z^\prime$ are weaker,
with $M_{Z^\prime} \sim 150$~GeV and $\theta \sim$ a few per cent allowed. 
For reviews, see Refs.~\cite{Langacker95B}. 

Perturbative string models with supergravity mediated SUSY breaking usually 
predict many extra $Z^\prime$ bosons and exotic states. Unlike the GUT 
scenarios, these models tend to favor $Z^\prime$ masses of ${\cal O} (M_Z)$. 
The general idea~\cite{Cvetic97} is that the mixing of the two MSSM Higgs 
doublets (the $\mu$ term) which is expected to vanish at tree level (and 
in fact to all orders in the absence of SUSY breaking) is generated 
radiatively through large Yukawa couplings. The $\mu$ problem, i.e., 
the expectation that either $\mu = 0$ or of order the Planck scale, is solved 
by the introduction of a SM singlet field $S$. Its expectation value generates
an effective $\mu$ term of ${\cal O} (M_H^2) \sim {\cal O} (M_S^2) \sim 
{\cal O} (M_{\rm SUSY}^2)$. Given this kind of scenario, a TeV scale 
$Z^\prime$ boson is no longer ``lamppost physics''.

Another class of models~\cite{Cleaver98A} involves symmetry breaking along 
D-flat directions. This can introduce $Z^\prime$ bosons with intermediate 
scale masses (e.g. $10^{12}$~GeV), and can have interesting consequences for 
the generation of fermion mass hierarchies and neutrino masses.

\section{Conclusions}
The precision data confirms the validity of the SM at the electroweak loop 
level (i.e., within a few per mille), and there is no compelling evidence for 
deviations. The determination of $m_t$ from electroweak loops is consistent 
with the kinetic mass measurement, and the low versus high energy $\alpha_s$ 
problem has disappeared. The evaluation of $\alpha (M_Z)$ is due to new 
perturbative and non-perturbative QCD treatments more precise than in the past.

A low Higgs mass is strongly favored by the data. $M_H$ is now much more
robust to changes in the data set. The precise range is rather sensitive to 
$\alpha (M_Z)$, but the upper limits are not. However, in the presence of 
non-standard contributions to the $S$ or $T$ parameters, no strong $M_H$ bounds
can be found. 

There are stringent constraints on parameters beyond the SM, such as
$\rho_0$, $S$, $T$, $U$, and others. This is a serious problem for models
of dynamical symmetry breaking, compositeness, and the like. Those constraints
are, however, consistent with gauge unification, and also with the MSSM
favoring its decoupling limit. Moreover, the low favored $M_H$ is in 
agreement with the expected mass range for the lightest neutral Higgs boson 
in the MSSM.

Perturbative string models (and also many GUT's) suggest extra $Z^\prime$
bosons. Rather generically they are expected at the electroweak scale and are 
highly predictive, with many specific models already excluded. In other words,
a TeV scale $Z^\prime$ is no longer ``lamppost physics'', but is among the best
motivated possibilities beyond the MSSM. In other scenarios extra $Z^\prime$ 
bosons can appear at intermediate scales, and could solve the fermion mass 
hierarchy problem, and simultaneously generate neutrino masses. 


\section*{Acknowledgments}
We would like to thank Bob Clare for providing us with details of the
LEP lineshape and LEP/SLC heavy flavor fits, and Wolfgang Hollik and
Hans K\"uhn for discussions.

\end{document}